\documentclass[3p,times]{elsarticle}
\usepackage{ecrcpub}

\volume{824 (2022)}

\firstpage{136801}

\journalname{Physics Letters B}

\runauth{S.J. Brodsky, V.E. Lyubovitskij, I. Schmidt}

\jid{plb}

\jnltitlelogo{Physics Letters B}

\usepackage{amssymb,amsmath,amsfonts}
\usepackage{epsfig,graphicx,tabularx}
\usepackage{color}
\usepackage[figuresright]{rotating}

\newcommand{\seq}{\begin{subequations}}
\newcommand{\sen}{\end{subequations}}
\newcommand{\eq}{\begin{eqnarray}}
\newcommand{\en}{\end{eqnarray}}
\newcommand{\ra}{\rangle}
\newcommand{\la}{\langle}

\begin{document}

\begin{frontmatter}

\dochead{}

\title{The Diffractive Contribution to Deep Inelastic Lepton-Proton Scattering: 
Implications for QCD Momentum Sum Rules and Parton Distributions}  

\author[label1]{Stanley J. Brodsky} 
\ead{sjbth@slac.stanford.edu} 
\author[label2,label3,label4]{Valery E. Lyubovitskij\corref{cor}}
\ead{valeri.lyubovitskij@uni-tuebingen.de}
\author[label3]{Ivan Schmidt}
\ead{ivan.schmidt@usm.cl}
\address[label1]{SLAC National Accelerator Laboratory, 
Stanford University, Stanford, CA 94309, USA}
\address[label2]{Institut f\"ur Theoretische Physik,Universit\"at T\"ubingen,
Kepler Center for Astro and Particle Physics, \\
Auf der Morgenstelle 14, D-72076 T\"ubingen, Germany}
\address[label3]{Departamento de F\'\i sica y Centro Cient\'\i fico
Tecnol\'ogico de Valpara\'\i so-CCTVal, \\ Universidad T\'ecnica
Federico Santa Mar\'\i a, Casilla 110-V, Valpara\'\i so, Chile}
\address[label4]{Millennium Institute for Subatomic Physics 
at the High-Energy Frontier (SAPHIR) of ANID, \\
Fern\'andez Concha 700, Santiago, Chile} 
\cortext[cor]{Corresponding author}

\begin{abstract}

The cross section for deep inelastic lepton-proton scattering (DIS) $\ell p \to \ell' X$ includes 
a diffractive deep inelastic (DDIS) contribution $\ell p \to \ell' p' X$,  
in which the proton remains intact with a large longitudinal momentum fraction $x_F$ greater 
than 0.9 and small transverse momentum. The DDIS events, which can be identified with 
Pomeron exchange in the $t$-channel, account for 
approximately $10\%$ of all of the DIS events.  Thus, when one measures DIS,
one automatically includes the leading-twist Bjorken-scaling DDIS events as 
a contribution to the DIS cross section, whether or not
the final-state proton $p'$ is detected. 
In such events, the missing momentum fraction $x_{p'} \sim
0.9$ carried by the final-state proton $p'$ in the DDIS events could be misidentified with
the light-front momentum fraction carried by sea quarks or gluons in the protons' Fock structure. 
As we shall show in this article, the underlying QCD Pomeron-exchange amplitude which produces 
the DDIS events does not obey the operator product expansion nor satisfy momentum sum rules. 
Thus we conclude that the quark and gluon distributions measured in DIS
experiments will be misidentified, unless the measurements explicitly exclude the DDIS events and that a
correct determination of the parton distribution functions (PDFs) derived from the DIS data requires 
the explicit subtraction of the DDIS contribution from the full DIS cross section. 

\end{abstract}


\end{frontmatter}

\section{Diffractive Deep Inelastic Scattering}

Deep inelastic lepton scattering (DIS) $\ell + p \to \ell' + X$ ~\cite{Bloom:1969kc,Breidenbach:1969kd} 
on the proton is the key experimental tool for extracting information about the fundamental quark and 
gluon structure of the proton, as encoded in its partonic distribution functions (PDFs).   

A striking feature of DIS lepton-proton scattering is the large fraction 
of diffractive events $\ell + p \to \ell' + p' +X$, where the struck proton remains intact, 
with longitudinal Feynman momentum fraction $x_F$ greater than 0.9,
acquiring small momentum transfer, and being well-separated in rapidity from the other produced particles. 
As first discovered by the ZEUS Collaboration and 
rediscovered by the H1 Collaboration at DESY~\cite{Derrick:1993xh},
approximately 10\% of the conventional DIS events are diffractive. 
These diffractive deep inelastic scattering (DDIS) events are also observed to be leading twist; 
i.e., they satisfy Bjorken scaling [see Fig.4 in Ref.~\cite{Derrick:1993xh}]. 

Thus when one measures the DIS cross section,
one automatically includes the DDIS contribution, even when
the final-state proton $p'$ is not detected nor measured.
In such events, the missing momentum fraction $x_{p'} \sim
0.9$ carried by the final-state proton $p'$ in the DDIS events could be misidentified with
the light-front (LF) momentum fraction carried by sea quarks or gluons in the target proton's Fock structure.
As we show in this article, the underlying QCD Pomeron-exchange mechanism which produces the DDIS events 
obeys leading-twist Bjorken scaling, but it does not obey the operator product expansion (OPE) nor 
satisfy momentum sum rules (MSRs). Therefore, the PDFs measured in DIS
experiments will be misidentified if the DIS measurements do not exclude the DDIS events; the
true extraction of PDFs from the data requires 
the explicit identification and subtraction of the DDIS contribution from the full DIS cross section. 
For a comprehensive discussion of diffractive Pomeron contributions and other nonlinear contributions 
to the sea quark and gluon distributions measured in DIS, see Ref.~\cite{Watt:2005iu}.

We shall begin with a simplified parton-model interpretation of DDIS $e p \to e' p' X$, as measured 
in an $ep$ collider.  
For example, the electron can scatter on a sea quark $q$ in an extrinsic five-quark Fock state 
$|(uud)_{8C} (Q \bar Q)_{8C}>$ of the proton, a Fock state created from an internal virtual gluon exchange. 
The $e Q  \to e' Q'$  collision
produces a quark jet with high transverse momentum opposite the transverse momentum of the scattered electron,
while the $\bar Q$ and the $(uud)_{8C}$ remain as spectators with momenta similar to that as the incident proton.
Subsequently, after the electron-quark scattering $e Q \to e' Q'$ occurs, a final-state gluon can then be
exchanged between the $\bar q$ and one of the $(uud)_{8C}$ valence quarks.  
This final-state soft gluon exchange neutralizes the color of the ${uud}$ spectators to a $(uud)_{1C}$ 
color-singlet Fock state which can in turn overlaps with the eigenstate of a final-state proton $p'$. 
The produced $p'$  will have a 4-momentum closely matching the initial proton's 4-momentum. 

In this simple picture, two gluons are exchanged in the 
$t-$ channel between the $\gamma^* \to Q \bar Q$ and $p \to p'$ systems, as in 
the Low-Nussinov model~\cite{Low:1975sv,Nussinov:1975mw}. 
The final state of the DDIS events is thus characterized by a $q + \bar q$ dijet, 
plus a final state $p'$ close in rapidity and momentum to the rapidity and momentum of the initial proton. 
Since the second final state gluon exchange happens after the electron scatters, 
the scattering amplitude has a propagating intermediate state with a phase $i$ corresponding 
to a Glauber cut. Thus DDIS is in effect Pomeron exchange.
Since they are diffractive, the DDIS events can be characterized as due 
to soft Pomeron exchange between the sea quark pair, which is produced by the virtual photon 
$\gamma^*(q^2) \to Q\bar Q$ and the target proton. 
Therefore, in this Pomeron based-picture, 
the  DDIS amplitude  $\gamma^* + p  \to Q \bar Q + p'$ has the energy and phase 
dependence $M(s,t,q^2) \sim i s \beta(t)$ of Pomeron exchange. 
One should stress that our analysis also applies for scattering amplitudes
where the Pomeron exchange corresponds to two hard gluons. 

One can also use Regge phenomenology 
and extend the class of diffractive events to $C=-$ Odderon exchange, as well as Reggeon 
quark interchange events with isospin exchange, such as $\gamma^* + p \to n + X^+$.  
In addition, the final state $p'$  
in DDIS events can also be an excited proton state or a nucleon resonance. 
DDIS events can also occur in lepton-ion collisions $\gamma^* + A \to A + X$, 
where the nucleus $A$ remains intact in its ground state.

It is also possible to still tag the flavor of the struck sea quark $Q'$ in DDIS events, 
by measuring the leading 
hadron in the jet recoiling opposite to the lepton. For example, one can tag strange
quark events by measuring a leading strange hadron at high $z$ in the jet; this
measures the strange quark distribution $s(x,Q^2)$ or $\bar s(x,Q^2)$ in the proton
where its LF momentum $x=x_{Bj}$ is determined by the lepton kinematics.

\section{Light-Front Description of DDIS}

How can the distinction between the conventional DIS  $\gamma^* p \to X$ and DDIS events  
$\gamma^*  p \to p' X$ be understood in terms of the underlying quark and gluon degrees  
of freedom of QCD ?  
In the standard, frame-independent description based on the light-front (LF) Hamiltonian 
formulation of QCD, the DIS $\gamma^* p \to X$ events are due to the interaction of 
a virtual photon $\gamma^*(q)$, emitted by the incident lepton scattering on a quark or antiquark on 
one of the Fock states of the proton eigenstate $
H^{\rm QCD_{LF}} |\Psi_p\rangle = M^2_p |\Psi_p\rangle$ of QCD, 
via the free quark current $j^+(0)$. 
The standard LF kinematics for DIS are:
$p^\mu = (P^+, \vec 0_\perp, \frac{M^2}{P^+})$, 
$q^\mu = (0, \vec q_\perp, q^-)$, 
$\vec q_\perp^2 = Q^2 =-q^2$.
Therefore, $(q+p)^2 = q^-P^+= W^2$ and $q^- = \frac{W^2}{P^+}$. 
The quark and gluon constituents in each Fock state of the proton have kinematics 
$k^\mu =  (x p^+, \vec k_\perp, \frac{m^2+k^2_\perp}{x}) $,
where $\sum_i x_i = 1$ and  $\sum_i \vec k_{\perp i} = \vec 0_\perp$. Notice that this LF formulation 
is invariant under both longitudinal and transverse boosts, 
and that the struck quark or antiquark has 
the LF momentum fraction $x = x_{Bj} = \frac{Q^2}{2p\cdot Q}$.  

The probability amplitude for a given Fock state $|n\rangle$ in the target proton is given by 
its light-front wavefunction (LFWF) 
$\phi^p_n (x_i, {\vec k}_{\perp i}, \lambda_i) $,  which is 
the projection  $\langle n | \Psi_P \rangle$  of the proton eigenstate on the quark and 
gluon Fock state $|n\rangle $ of the free Hamiltonian.  
The structure functions, transverse momentum distributions and their evolution measured in DIS  
can thus be predicted from the square of the  LFWFs, summed over the contributing Fock states.  

The kinematics of the lepton-proton collision is chosen such that the virtual photon $\gamma^*(q^\mu)$ 
has $q^+=0$.  This choice of kinematics eliminates events where the virtual photon creates 
a $Q \bar Q$ pair in the lepton-proton collision, since the LF momentum $k^+$ 
of each particle in LF Hamiltonian theory is positive, and the $+$ momentum is conserved. 
The virtual photon has therefore no quantum fluctuations and acts as a classical probe of the proton's 
structure.  For example, extrinsic heavy quark production in lepton-proton collisions is solely associated 
with the proton's structure.
(This is the LF equivalent of  Feynman's choice of the infinite momentum frame $P^z\to \infty$ 
in ordinary instant-form Hamiltonian theory.) The LF formulation thus provides a rigorous 
foundation for the parton model. 

As noted above, the DDIS $\gamma^* p \to p' X$ events begin with the $\gamma^*$ interacting with a 
non-valence quark $Q$ or $\bar Q$ in any LF extrinsic Fock state of the proton, which contains sea quarks, 
such as $|(uud )_{8C} (Q\bar Q)_{8C}\rangle$.  Such an five-particle extrinsic Fock state 
can originate from a virtual color-octet gluon creating the color octet $Q \bar Q$ pair. 

\begin{figure}[ht!]
\includegraphics[scale=0.2]{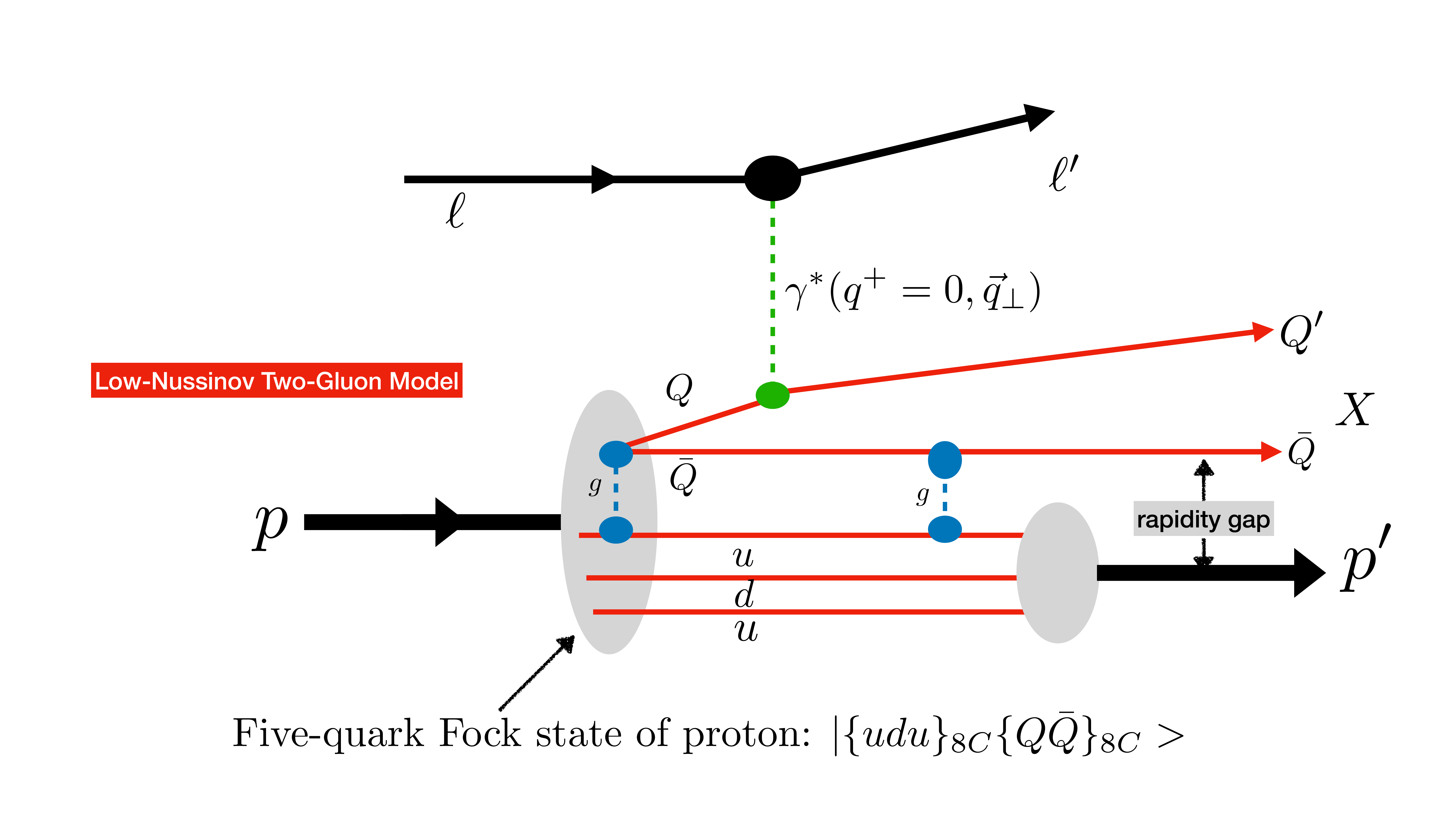}
\vspace*{-.75cm}
        \caption{Simplified description of the DDIS event
$\gamma^* + p_{|uduQ\bar Q\rangle} \to p' + X + {\rm rapgap}$
from the simplified two-gluon Low-Nussinov model of Pomeron exchange in the LF framework.   
Multiple gluons can also propagate between the initial and final gluons.
The five-quark Fock state of proton $|\{udu\}_{8C}\{Q\bar Q\}_{8C}\rangle$ 
produces the rapidity gap.}
\label{fig1}
\end{figure}

We can now follow the evolution of the event in LF time $\tau = t+z/c$, after the 
$\ell Q \to \ell' Q'$ interaction.  See Fig.~\ref{fig1}. 
The struck quark $Q'$ acquires large transverse momentum $\vec q_\perp$, with kinematics 
$\Big(xP^+, \vec q_\perp, \frac{m^2_Q + {\vec q_\perp}^2}{xP^+}\Big)$.
The $Q'$ and its partner $\bar Q$  then propagate as a massive
$(Q' \bar Q)_{8C}$ color octet, and the remaining spectator valence quarks $(uud)_{8C}$ propagates 
as a color-octet along the proton direction. 
As the LF time $\tau$ progresses, a soft gluon exchange can be exchanged between the $\bar Q$ of the 
propagating  $(Q' \bar Q)_{8C}$ and the $(uud)_{8C}$. This produces  
a color-singlet $(Q \bar Q)_{1C}$  dijet system and a final-state color-singlet cluster $(uud)_{1C}$ 
which can materialize as the final state proton $|p'\rangle$ 
thus giving the observed $\gamma^* p \to p' X$ DDIS event.  The final state $X$ can thus be 
a $Q\bar Q$ dijet isolated in rapidity from the remnant forward moving $p'$.

The DDIS event is leading-twist, 
Bjorken-scaling in $q^2$, since the only hard interaction occurring is the hard scattering 
of the lepton on the $Q$ of the initial proton  $|uud Q\bar Q\rangle$ Fock state. 
This DDIS event mimics the exchange of the Low-Nussinov model in the $t$ channel.  
The initial exchanged soft gluon in the proton Fock state produces the
$|(uud )_{8C} (Q\bar Q)_{8C}\rangle$ configuration in the initial state; 
the final-state exchange of the second soft gluon produces the isolated color singlet $Q\bar Q$,
and the final-state color singlet $|uud\rangle$ produces the intact proton $p'$ with small 
momentum transfer $t$. One can also have multiple gluons propagating between the initial and final 
gluons, thus resembling a model of the Pomeron 
with the exchange of two interacting gluon strings in the $t$ channel.

\section{Application of the OPE to DIS}

\begin{figure}[ht!]
\includegraphics[scale=0.2]{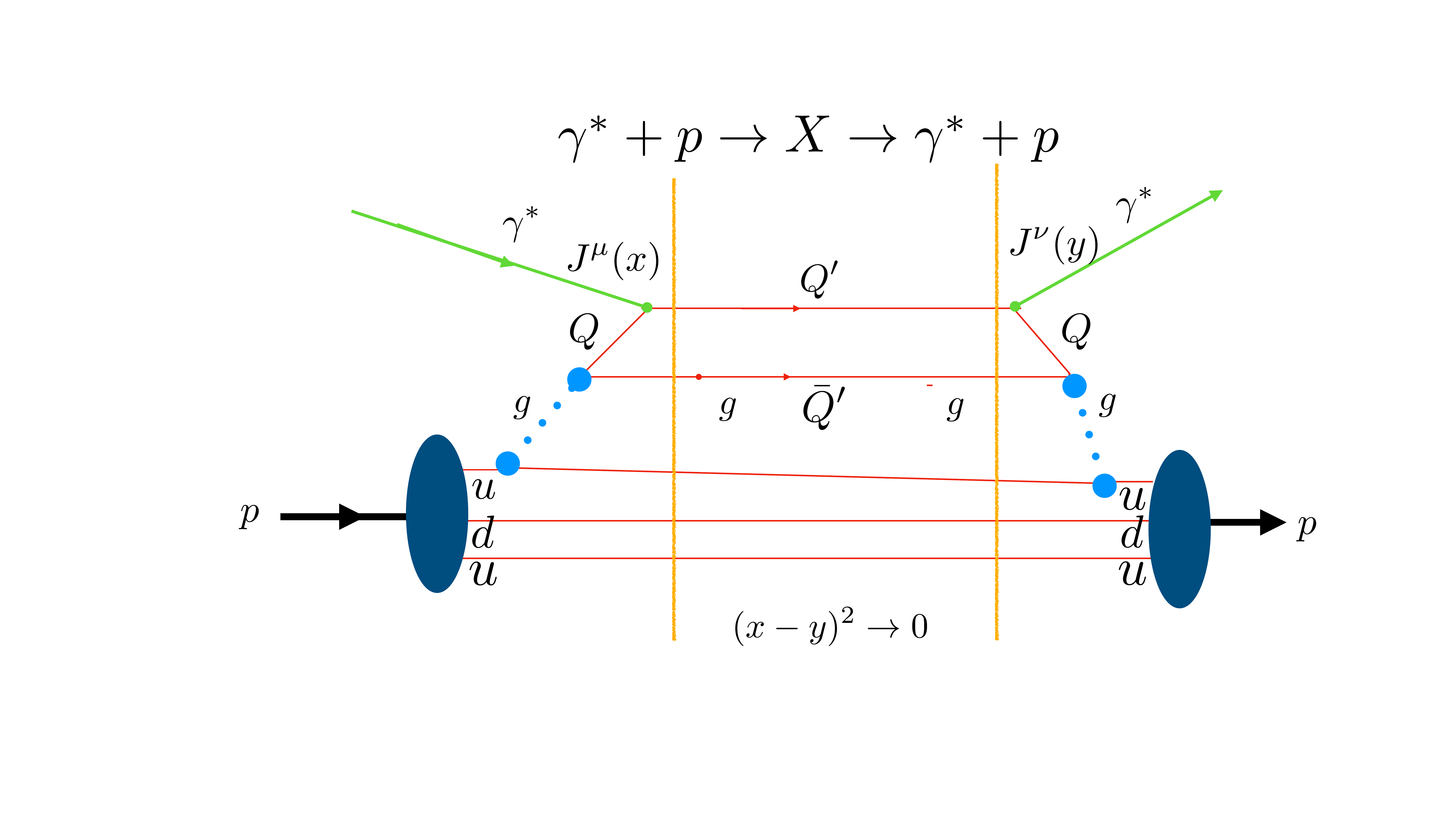}
\vspace*{-1.45cm}
        \caption{Forward DVCS for a DIS event $\gamma^* + p \to X \to \gamma^* + p$. 
Vanishing LF time between currents of virtual photons at high $q^2$ allows the 
application of the OPE to DIS.} 
\label{fig2}
\end{figure}

The application of the OPE to DIS utilizes the properties of deep inelastic 
forward virtual Compton scattering (DVCS) $\gamma^* p \to \gamma^* p$, since
the imaginary part of its amplitude gives the DIS cross section $\gamma^* p \to X$ by unitarity.
In the case of a simple DIS reaction $\gamma^* + Q \to Q'$, where the lepton scatters on a quark $Q$, 
the DVCS amplitude has the form of the {\it handbag diagram}, since 
there are no interactions occurring in LF time which will interrupt the propagation of the struck 
quark $Q$. See Fig.~\ref{fig2}.

In DIS at large $q^2$, the  amplitude $\gamma^* + Q \to \gamma^* + Q$ contracts to an effective 
seagull operator of the local product of quark currents $j^+(0) j^+(0)$, which in the OPE can be 
replaced by the gravitational energy-momentum tensor $T^{++}(0)$. The graviton coupling is 
proportional to the LF momentum fraction $x =\frac{k^+}{P^+}$ carried by the struck quark, while
the resulting matrix element $\langle p|T^{++}(0)|p\rangle$  is given by square of the proton's LFWF 
integrated over the quark, antiquark and gluon $x_i$ and $\vec k_{\perp i}$ kinematics.
In Fig.~\ref{fig2}, the sequence in LF time is marked by the vertical orange bars. 
This procedure underlies the OPE derivation of the MSR for structure functions.
One thus obtains the total LF momentum carried by the charged quarks in the proton. 

The MSR is a property of the LFWF of the target proton; i.e., it is a property of the eigenfunction 
of the QCD light-front Hamiltonian. The square of the proton's LFWF defines the proton structure 
function at an initial scale, which then evolves by DGLAP evolution. The momentum sum rule is 
maintained by the pQCD subprocesses described by DGLAP evolution such as $q \to q g$ and $g \to q \bar q$. 
However, the diffractive contribution due to soft gluon exchange in the final state, 
which keeps the proton intact at low momentum transfer, is not due to the subprocesses 
which underly DGLAP evolution. Moreover, the DDIS amplitude acquires the complex phase of 
$t$-channel Pomeron exchange, physics which is not described by the usual leading-twist pQCD contributions. 
The diffractive DIS contribution from Pomeron exchange is thus a distinct contribution 
to the DIS cross section which does not obey the momentum and other sum rules.   
Note also that diffractive processes underlie the shadowing and anti-shadowing of the 
nuclear cross sections. Thus, as discussed in Ref.~\cite{Nussinov:1975mw},  
nuclear parton distribution functions are not constrained by the momentum or other sum rules. 
Note that the measurements by the H1 Collaboration [4] for the diffractive F2 structure function
satisfied the MSR within uncertainties; the MSR was not imposed in the fit.
However, the NuTeV measurement of charged current DDIS contradicts the expectation that 
anti-shadowing compensates shadowing to restore the MSR for nuclear structure functions.

\begin{figure}[ht!]
\includegraphics[scale=0.2]{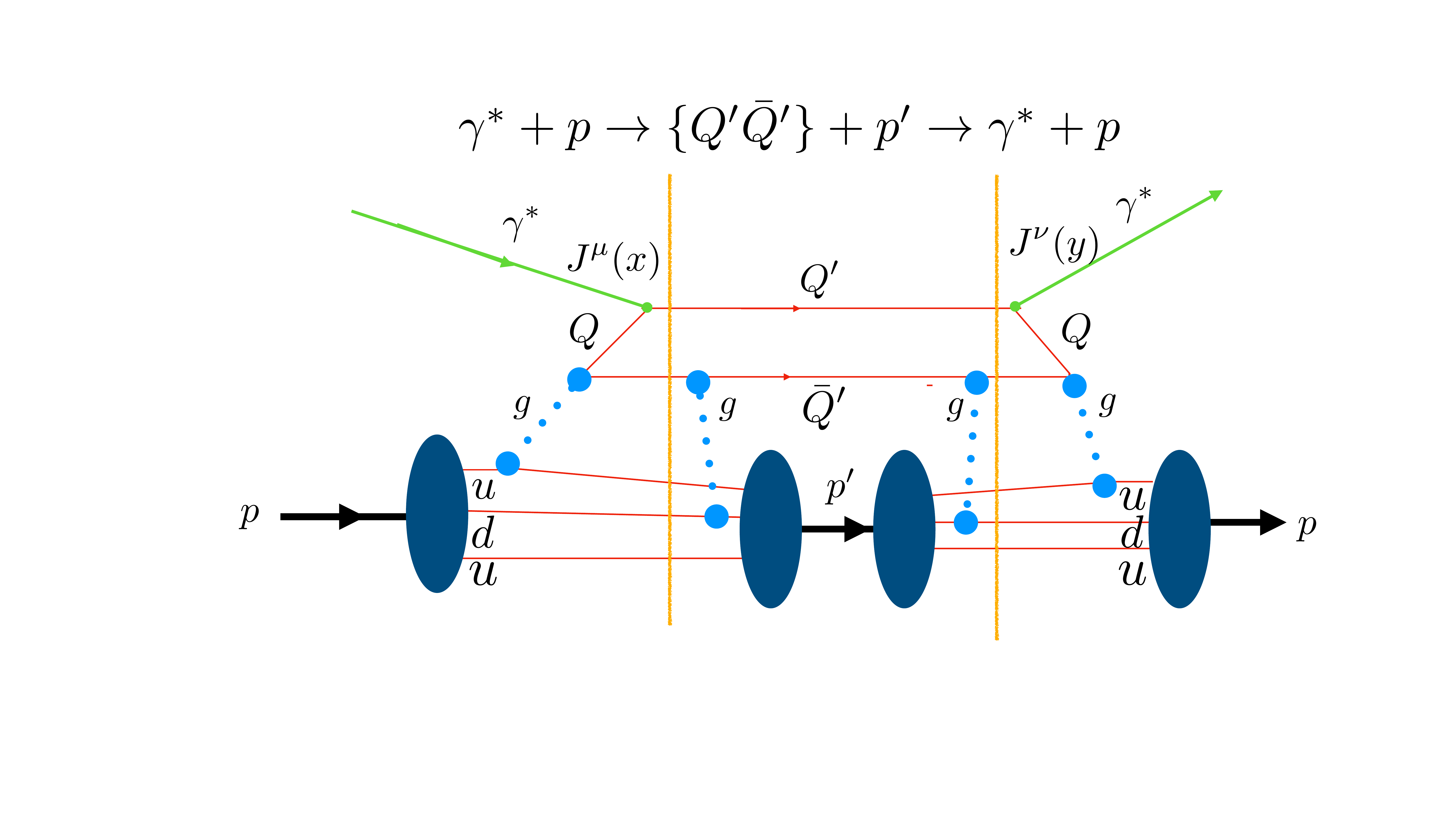}
\vspace*{-1.45cm}
        \caption{Forward DVCS for a DDIS event. 
Nonzero LF propagation time between virtual photons cannot be reduced 
to a local seagull interaction, thus preventing the application of OPE.} 
\label{fig3}
\end{figure}

Fig.~\ref{fig3} shows the corresponding contribution to 
forward DVCS $\gamma^*  + p \to  \{Q \bar Q\}  + p' \to  \gamma^* + p$  
from the DDIS channel.
In this case, the imaginary part of the forward DVCS amplitude gives the DDIS cross section 
and thus its contribution to the DDIS structure function.
The sequence in LF time is again marked by the vertical orange bars.
However,  because of the final-state interaction (FSI) of the virtual gluon between the struck  
sea quark $Q$ and the valence quarks, which restores the proton intact in its ground state in DDIS, 
the LF time $\Delta \tau = x^+ - y^+$ between the currents $J^\mu(x)$  and $J^\nu(y)$  in the DVCS 
amplitude is interrupted.  
The interactions of the two currents on the quark propagator in the DVCS amplitude
cannot be reduced to a local seagull interaction, thus  preventing the usual application of the OPE. 
We stress that the forward DVCS process shown in Fig.~\ref{fig3}
represents the standard leading-twist two-gluon Low-Nussinov
contribution to DDIS, where the proton stays intact and isolated.
The extra $1/Q^2$ factors in the forward virtual Compton amplitude
from the two extra denominators are compensated by the gluon couplings.

Our analysis is also supported by the results of Ref.~\cite{Pelicer:2018xzu},
where the importance of absorptive correction at low $x$ are noted.
These authors use the knowledge of diffractive deep inelastic scattering
to exclude the absorptive effect from conventional DIS data.
Gluon recombination at low $x$ leading to saturation of the gluon density
is a rather well known phenomena. The authors of Ref.~\cite{Pelicer:2018xzu}
use the known diffractive part of the DIS cross section `` to remove
the absorptive corrections from the inclusive data. 
The MSR is therefore inapplicable for the fraction of DIS events 
which are leading-twist DDIS events.

A key point is that only the final-state quarks which are scattered out with high
transverse momentum opposite to the final-state lepton are measured in the DIS
measurement. In the standard parton model picture, one assumes that all (100\%) of
the momentum of the incident proton in DIS events is carried off in the final state by
quarks, antiquarks, and gluons. The MSR $\sum\limits_{i=q,\bar q, g}
x_i =1$ is then used to obtain the gluon momentum fraction $\la x_g \ra$. 

If one sums the contributions $\sum_i x_{q i}$ to the LF MSR from the outgoing
scattered quarks, the LF momentum fraction $x_p' \sim 0.9 $ which is carried off by $p'$
in the DDIS events is not counted. In this case, the spectators include the final-state intact proton;
the scattered quarks thus cannot give accurate information about the parton momentum
distribution inside the proton. Then substantial LF momentum $x_p' $ which is carried off
by $p'$ in DDIS events is misinterpreted as the gluon LF momentum fraction $\la x_g \ra$.
The value for $\la x_g \ra$ -- which has been obtained by misapplying the MSR -- has thus
been overestimated -- perhaps by $ 0.15 \times 0.9 \sim 0.13$. This overestimate of
$\la x_g \ra$ can also have logarithmic dependence on $Q^2$. 

The FSI interactions of the included DDIS events from gluon exchange severely distort
the measured sea-quark and antiquark distributions. The DIS experimentalists will thus
obtain a false determination of $\la x_q \ra$ and $\la x_{\bar q} \ra$, because they 
incorrectly include the DDIS events, and when the MSR is applied, one obtains a false
determination of $\la x_g \ra$. The imaginary part (cut) gives the cross section for the DDIS
event $\gamma^* p \to p' x$, and then the LF time interval between the two currents at $x$
and $y$ does not vanish at large $Q^2$, since it includes the nonzero time for the proton
to propagate. One cannot relate the product of $J^\mu(x)$ and $J^\nu(y)$ to the local
operator $T^{\mu \nu}$. Thus the OPE is invalid and the MSR cannot be derived. 

\begin{figure}[ht!]
\includegraphics[scale=0.20]{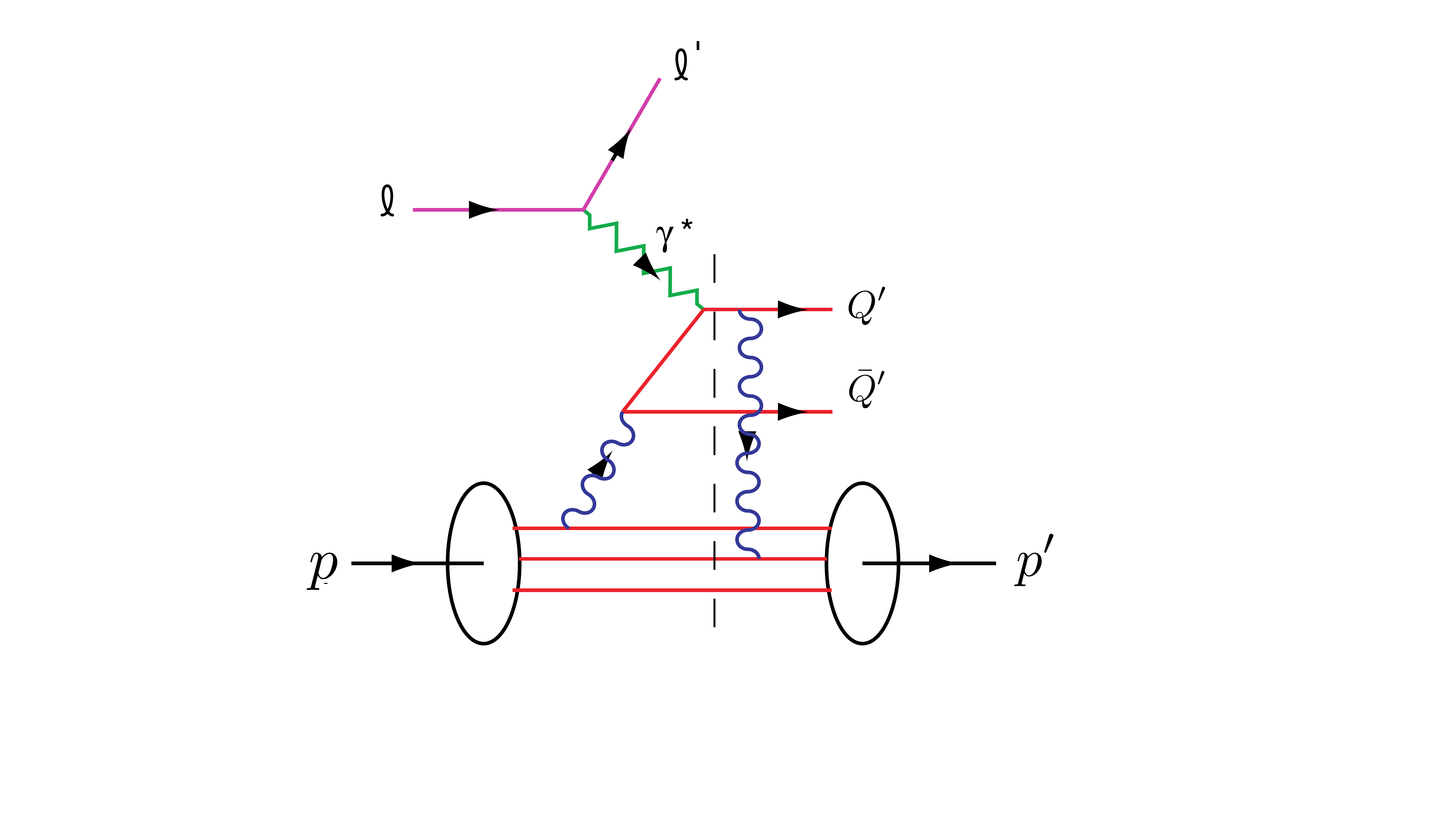}
\vspace*{-1.55cm}
        \caption{DDIS eventfinal-state interactions  $\gamma^* + p \to  p' + X$ 
with $Q'\bar Q'$ pair production. 90\% of proton momentum 
carried off by final proton $p'$ in 15\% of events. 
Gluon momentum fraction misidentified. 
$p'$ is produced in DDIS but 
escapes detection, and is misinterpreted as a DIS event.}
\label{fig4}
\end{figure}

The DDIS $\gamma^* p \to  p' X$ events where the final-state proton $p'$ is produced -- 
with a rapidity gap separating it from the other final-state hadrons -- are a substantial subset of 
the final states which contribute to the leading-twist $\gamma^* p \to  X$ DIS cross section.   
However, as we have shown, the DDIS events should be excluded when evaluating the traditional 
DIS momentum and other sum rules. In fact, as indicated in Fig.~\ref{fig4}, the DDIS events will be 
misinterpreted as contributions to the gluon momentum fraction.

The DDIS events also play a role in the theory of shadowing and antishadowing of nuclear cross sections 
and thus the application of the MSR to nuclei, as we have discussed 
in Ref.~\cite{Brodsky:2019jla}. 

The FSI occurring in DDIS phenomena has a similarity with the leading-twist 
Sivers' single-spin asymmetry (SSA), which is due to FSIs 
in semi-inclusive DIS~\cite{Brodsky:2002cx}, and which are not included 
in the proton's LFWFs. In particular, SSA does not factorize into 
a product of PDF and fragmentation function, and 
it is not related to the quark transversity PDF. 
As noted above, the DDIS cross section is also leading twist -- Bjorken scaling. 
The numerator of the final-state gluon exchange amplitude compensates for the $1/Q^2$. 
In fact, the same gluon exchange interaction gives the leading-twist Sivers effect. 
The imaginary phase comes from the Glauber cut and contributes to both DDIS and the Sivers effect. 
In particular, the SSA does not factorize into 
a product of PDF and fragmentation function, and 
it is not related to the quark transversity PDF. Another
example is the double Boer-Mulders effect in DY which breaks the Lam-Tung relation.

Thus the DDIS events do not actually measure the physics of the LFWF of the
target proton since they arise from FSIs which are not in the
eigensolution of the QCD LF Hamiltonian. The LFWF is a property of the hadronic
eigenstate of the QCD LF Hamiltonian, and it obeys the MSR $\sum\limits_{i=q,\bar q, g}
\la x_i \ra =1$. The LFWF has no knowledge of events that occur after the electron quark
scattering.

\section{The ``True", Non-Diffractive DIS Cross Section.}  

Now we turn to the master formulas which define the ``true" non-diffractive DIS cross section (NDDIS)
and true PDFs without DDIS events. The double-differential DIS cross section depending 
of $x$ and $Q^2$ variables is expressed in terms of structure functions 
$F_2$ and $F_L = F_2 - 2x F_1$ as 
\eq\label{DIS1}
\frac{d\sigma_{\rm DIS}}{dx dQ^2} = \frac{4\pi\alpha^2}{Q^4}
\, \frac{1+(1-y)^2}{x} \, 
\biggl[ F_2 - F_L \, \frac{y^2}{1+(1-y)^2} \biggr] \,, 
\en
where $Q^2, \nu = p \cdot q \to \infty$. 
DDIS cross section is defined by analogy 
\eq\label{DIS2}
 \frac{d\sigma_{\rm DDIS}}{dx dQ^2} = \frac{4\pi\alpha^2}{Q^4}
\, \frac{1+(1-y)^2}{x} \, 
\biggl[ F_2^D - F_L^D \, \frac{y^2}{1+(1-y)^2} \biggr] \,, 
\en
where $F_2^D$ and $F_L^D=F_2^D - 2x F_1^D$ are the diffractive structure 
functions. 

As we have shown the  DIS differential 
cross section which obeys the OPE and the MSR requires that the DDIS events to be
subtracted: 
\eq\label{DIS_true}
\frac{d\sigma_{\rm DIS_{true}}}{dx dQ^2} = 
\frac{d\sigma_{\rm DIS}}{dx dQ^2} -
\frac{d\sigma_{\rm DDIS}}{dx dQ^2} \,.
\en 
for each scattered quark $q'$.
Since the final states are different, this procedure reproduces the ``true'' quark
distributions $q(x,Q)$ and $\bar q(x,Q)$, which specifically come from the proton LFWF,
as in the parton model. Then the net result -- after subtracting the DDIS events --
is that the ``true'' sea quark and antiquark distributions obtained from 
$d\sigma_{\rm DIS_{true}}/dx dQ^2$ would appear at higher $x$. Applying the MSR for the partons in
the proton LFWF $\sum\limits_{i=q,\bar q, g} \,  \la x_i \ra  =1$ then gives a decreased value for the
``true'' gluon momentum fraction $\la x \ra_g$.

According to the factorization theorem, in QCD 
the DIS and DDIS structure functions are written as convolution 
of PDFs $f_i(x)$ (nonperturbative part) of flavor $i$ 
and hard scattering part (coefficient 
functions $C_{2,i}$)  
\eq
F_{2/L}(x,Q^2) = \sum\limits_{i} \, \int \frac{d\xi}{\xi} \, 
C_{2/L}^i\Big(\frac{x}{\xi}\Big) \, 
f_i(x,Q^2) 
\en 
for DIS 
and 
\eq 
F_{2/L}^D(\beta,Q^2;x_\mathbb{P},t) 
= \sum\limits_{i} \, \int \frac{d\xi}{\xi} \, 
C_{2/L}^i\Big(\frac{\beta}{\xi}\Big) \, 
f_i^D(\xi,Q^2;x_\mathbb{P},t) 
\en 
for DDIS, where $\beta$ is the diffractive exchange, 
$x_\mathbb{P} = x/\beta$ is the fraction of the momentum 
of proton carried by the diffractive exchange. 
Following Ref.~\cite{Vermaseren:2005qc,Maktoubian:2019ppi}, 
we use the same coefficient functions $C_{2/L,i}$ 
for DIS and DDIS. 
In the Regge factorization scheme separating 
the $Q^2$ and $x_\mathbb{P}$ behavior of the PDFs, 
DDIS PDFs are splited into the Pomeron $(\mathbb{P})$ 
and Reggeon $(\mathbb{R})$ parts as: 
\eq 
f_i^D(\beta,Q^2;x_\mathbb{P},t) = 
f_{\mathbb{P}/p}(x_\mathbb{P},t)\, 
f_{i/\mathbb{P}}(\beta,Q^2) \,+\,   
f_{\mathbb{R}/p}(x_\mathbb{P},t)\, 
f_{i/\mathbb{R}}(\beta,Q^2) \,, 
\en 
where $f_{\mathbb{P}/p}$ and $f_{\mathbb{R}/p}$ are 
the Pomeron and Reggeon fluxes, 
$f_{i/\mathbb{P}}(\beta,Q^2)$ and $f_{i/\mathbb{R}}(\beta,Q^2)$  
are the Pomeron and Reggeon PDFs.  
For $f_{\mathbb{P}/p}$ and $f_{\mathbb{R}/p}$ we use 
the same functional form~\cite{Maktoubian:2019ppi}: 
$f_{\mathbb{P},\mathbb{R}/p} = A_{\mathbb{P},\mathbb{R}} 
\, e^{B_{\mathbb{P},\mathbb{R}} t}/x_{\mathbb{P}}^{2\alpha_{\mathbb{P},\mathbb{R}}(t)-1}$,  
where $A_{\mathbb{P},\mathbb{R}}$ is the Pomeron/Reggeon 
intercepts. 
$\alpha_{\mathbb{P},\mathbb{R}}$ is the linear 
Pomeron/Reggeon trajectories: 
$\alpha_{\mathbb{P},\mathbb{R}}(t) = 
\alpha_{\mathbb{P},\mathbb{R}}(0) + 
\alpha_{\mathbb{P},\mathbb{R}}' t$. 
For the DDIS PDFs $f_{i/\mathbb{R}}(\beta,Q^2)$ we 
use parameterizations extracted from world data analysis, 
which for quarks/antiquarks and gluon read: 
$f_{i}(z,Q_0^2) = \alpha_i \, 
z^{\beta_i} \, (1-z)^{\gamma_i} \, (1 + \eta_i \sqrt{z})$\,,  
\ $i=q,g$. 
As commonly used, we assume that the quark and antiquark PDFs are equal. 
The sets of input parameters are listed, e.g., in~\cite{Maktoubian:2019ppi}. 

Upon the assumption on the same coefficient functions $C_{2/L,i}$ 
for DIS and DDIS, we can define the true PDF, 
as full PDF from which the diffractive PDF 
is subtracted: 
\eq\label{PDF_true} 
f_i(x,Q^2)\bigg|_{\rm true} = 
f_i(x,Q^2) - 
f_i^D(x,Q^2) \,, 
\en 
where  
$f_i^D(x,Q^2) = f_{\mathbb{P}/p}(x_\mathbb{P}(0),0)\, 
f_{i/\mathbb{P}}(\beta_0,Q^2) \,+\,   
f_{\mathbb{R}/p}(x_\mathbb{P}(0),0)\, 
f_{i/\mathbb{R}}(\beta_0,Q^2)$, 
$f_{\mathbb{P},\mathbb{R}/p}(x_\mathbb{P}(0),0) =  
A_{\mathbb{P},\mathbb{R}}/x_{\mathbb{P}}^{2\alpha_{\mathbb{P},\mathbb{R}}(0)-1}$, 
and $\beta_0 = x/x_\mathbb{P}(0)$. 
DDIS PDFs must be manifestly independent on $x_\mathbb{P}(0)$ 
[or $\beta_0 =  x/x_\mathbb{P}(0)$ if they are expressed through $\beta_0$ instead of 
$x_\mathbb{P}(0)$] with taking into account of 
the constraint on a choice of the parametrization for DDIS PDFs: 
\eq 
\frac{\partial f_i^D(x,Q^2)}{\partial x_\mathbb{P}(0)} = 
\frac{\partial}{\partial x_\mathbb{P}(0)} \, 
\biggl[ f_{\mathbb{P}/p}(x_\mathbb{P}(0),0)\,                                                               
f_{i/\mathbb{P}}(x/x_\mathbb{P}(0),Q^2) \,+\,
f_{\mathbb{R}/p}(x_\mathbb{P}(0),0)\,                                                                           
f_{i/\mathbb{R}}(x/x_\mathbb{P}(0),Q^2) \biggr] = 0 \,. 
\en 
Therefore, the idea of extraction of realistic quark PDFs in proton from 
true DIS is simply reduced to the subtraction of PDF itself, following 
Eq.~(\ref{PDF_true}). 

In conclusion, we propose a definition of the true NDDIS, 
which yields the true PDFs by subtracting the respective DDIS contribution. The DDIS contribution to 
the total DIS is not negligible, and therefore, its separation and subtraction is required in order to 
obtain an accurate determination of the quark and gluon PDFs, which play important role in the description 
of hadron structure. 
Our paper provides new and important perspectives on the physics origin of
the entire class of Bjorken-scaling diffractive events in deep inelastic
lepton-proton scattering. Our analysis also has important implications
for physics underlying the shadowing and antishadowing of nuclear structure 
functions. This point was discussed in Ref.~\cite{Brodsky:2019jla}. 

Our main observation is that the quark and gluon parton distributions intrinsic to hadron structure
will be misidentified, unless one excludes the DDIS events;  for example, the correct determination 
of the PDFs for proton derived from the DIS data $\gamma^* p \to X$ requires the explicit subtraction 
of the leading-twist DDIS contribution $\gamma^* p \to  p' X$ from the full DIS cross section.  
In particular, the light-front momentum carried by gluons in the proton will be misidentified since 
the final-state proton $p'$ carries off LF momentum in the DDIS events.
The modification of PDFs is expected to be of order 10\%. This will also affect predictions for 
hadron production processes based on the standard factorization formalism.
Clearly, a careful separation of non-diffractive and diffractive
contributions to the measured deep inelastic cross section, as well
as other hard-process cross sections is required.

\section*{Acknowledgments}

This work was funded by BMBF
``Verbundprojekt 05P2018 - Ausbau von ALICE                                                                       
am LHC: Jets und partonische Struktur von Kernen''
(F\"orderkennzeichen No. 05P18VTCA1), by ANID (Chile) under
Grant No. 7912010025,  by ANID PIA/APOYO AFB180002 (Chile),
by FONDECYT (Chile) under Grants No. 1191103 and No. 1180232,
by Millennium Institute for Subatomic Physics
at the High-Energy Frontier (SAPHIR) of ANID, Code: ICN2019\_044.
The work of SJB was supported in part by the Department of Energy
under contract DE-AC02-76SF00515. SLAC-PUB-17599.


\begin{thebibliography}{999}

\bibitem{Bloom:1969kc} 
  E.~D.~Bloom {\it et al.},
  Phys.\ Rev.\ Lett.\  {\bf 23}, 930 (1969).

\bibitem{Breidenbach:1969kd} 
  M.~Breidenbach {\it et al.},
  Phys.\ Rev.\ Lett.\  {\bf 23}, 935 (1969).

\bibitem{Derrick:1993xh}
  M.~Derrick {\it et al.} (ZEUS Collaboration),
  Phys.\ Lett.\ B {\bf 315}, 481 (1993).

\bibitem{H1:1995cha}
  T.~Ahmed \textit{et al.} (H1 Collaboration),
  Phys.\ Lett.\ B {\bf 348}, 681 (1995). 

\bibitem{Watt:2005iu}
  G.~Watt, A.~D.~Martin, and M.~G.~Ryskin,
  Phys.\ Lett.\ B {\bf 627}, 97 (2005). 

\bibitem{Low:1975sv} 
  F.~E.~Low,
  Phys.\ Rev.\ D {\bf 12}, 163 (1975).

\bibitem{Nussinov:1975mw} 
  S.~Nussinov,
  Phys.\ Rev.\ Lett.\  {\bf 34}, 1286 (1975).

\bibitem{Pelicer:2018xzu} 
  M.~R.~Pelicer, E.~G.~de Oliveira, A.~D.~Martin, and M.~G.~Ryskin,
  Eur.\ Phys.\ J.\ C {\bf 79},  9 (2019). 

\bibitem{Brodsky:2019jla} 
  S.~J.~Brodsky, I.~Schmidt, and S.~Liuti,
  arXiv:1908.06317 [hep-ph].

\bibitem{Brodsky:2002cx} 
  S.~J.~Brodsky, D.~S.~Hwang, and I.~Schmidt,
  Phys.\ Lett.\ B {\bf 530}, 99 (2002). 

\bibitem{Vermaseren:2005qc} 
  J.~A.~M.~Vermaseren, A.~Vogt, and S.~Moch,
  Nucl.\ Phys.\ B {\bf 724}, 3 (2005). 

\bibitem{Maktoubian:2019ppi} 
  A.~Maktoubian, H.~Mehraban, H.~Khanpour, and M.~Goharipour,
  Phys.\ Rev.\ D {\bf 100}, 054020 (2019).

\end{thebibliography}
\end{document}